\documentclass[11pt]{article}
\usepackage{moriond,epsfig}

\def\Journal#1#2#3#4{{#1} {\bf #2}, #3 (#4)}

\def\PRL{\em Phys. Rev. Lett.}
\def\PRD{{\em Phys. Rev.} D}

\def\be{\begin{equation}}
\def\ee{\end{equation}}
\def\bea{\begin{eqnarray}}
\def\eea{\end{eqnarray}}

\begin{document}
\vspace*{4cm}
\title{QCD Results from the CDF Experiment at $\sqrt{s}$=1.96 TeV}

\author{Michele Gallinaro}

\address{Laboratory of Experimental High Energy Physics, \\ The Rockefeller University, \\ 
1230 York Avenue, New York, NY 10021, USA \\
michgall@fnal.gov \\ (representing the CDF collaboration)
}

\maketitle\abstracts{
First QCD results obtained from the CDF experiment using Run II data are reported.
The Run~II physics program at the Tevatron started in the spring of 2001, 
with protons and anti-protons colliding at an energy of $\sqrt{s} = 1.96$~TeV.
The size of the data sample already compares to that of Run~I.
Results presented here include the measurement of the inclusive jet cross section, 
a search for new particles decaying to dijets, and a study of diffractive dijet events.
}

\section{Introduction}

Hadronic jets are one of the key signatures in hadron collider physics and, at the Tevatron collider,
probe the highest momentum transfer region currently accessible.
Since the start of Run~II the CDF experiment has collected millions of jet events.

The CDF detector has been upgraded during many years of dedicated efforts. 
The upgraded detector is almost all new.
Several improvements to the Run~I detector have been implemented including 
a new tracking volume, a more compact plug calorimeter, an improved muon coverage, and a system of forward 
detectors to allow studies up to very large rapidity regions.

Run~II proton-antiproton collisions started in March 2001 with a center-of-mass energy of 1.96~TeV,
approximately 10\% higher than in the previous Run~I.
After one year dedicated to its commissioning, the CDF detector has been in stable 
operation mode since February 2002 for most of its components.
Since then, data have been collected copiously and some of the analyses are reported here.

\section{Inclusive Jet Cross Section}

The measurement of the inclusive jet cross section provides a powerful test of perturbative QCD and
provides a sensitive probe of quark substructure down to a scale of $\approx10^{-17}$cm.
Run~I data exhibited an excess in jet cross section for events with high $E_T$ jets, when 
compared to the then current {\it Parton Distribution Functions} (PDFs).
The increase in center-of-mass energy from 1.80 to 1.96~TeV and the increased luminosity expected 
in Run~II will result in a larger kinematic range for measuring jet production.
With a data sample of comparable size to Run~I, jet production can already be 
measured at transverse energies far beyond those of Run~I.

Calorimeters are the elements of the detector used in this measurement and are 
segmented in pseudo-rapidity $\eta$ and azimuth $\phi$ with a projective tower geometry.
Jets are here
reconstructed in the central region ($0.1<|\eta|<0.7$) in order to be well measured.
Forward rapidity regions will be explored in
future analyses and can be used as a powerful discriminant to select new physics processes~\cite{cteq}.
Jet energies are reconstructed using the Run~I cone algorithm, with a radius $R=\sqrt{\Delta{\eta}^2+\Delta{\phi}^2}=0.7$.
The energy from the 
collision,
not associated with the hard scattering (underlying event energy),
is subtracted from the total jet energy.
The jet energy scale, which constitutes the largest source of systematic uncertainty (estimated to be 5\%),
was set by using the balancing of the photon energy with the jet $p_T$ in $\gamma$+jet events. 
Further understanding of the energy scale will reduce this uncertainty.

\begin{figure}\centering
\epsfig{figure=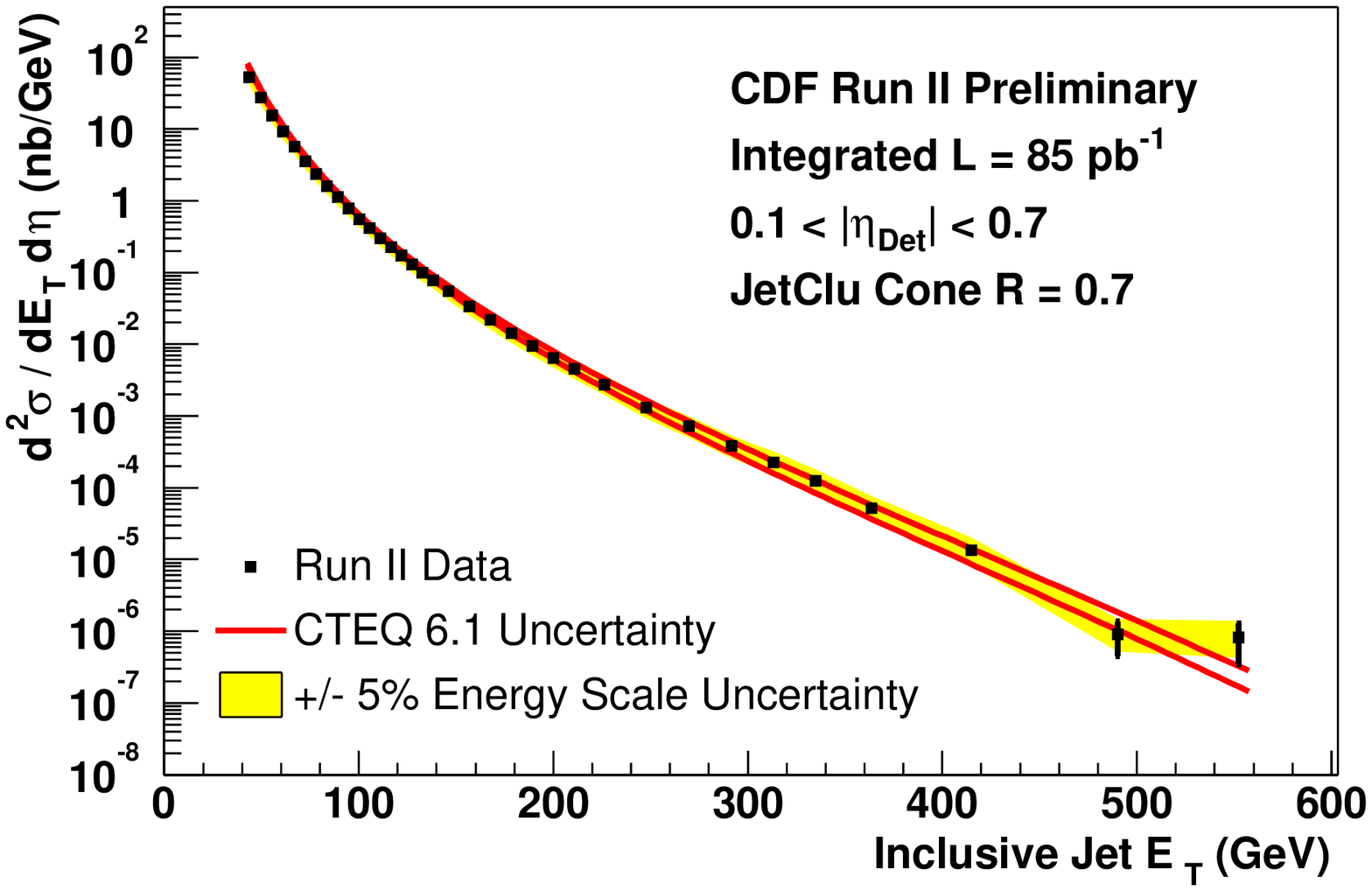,width=0.49\hsize}
\epsfig{figure=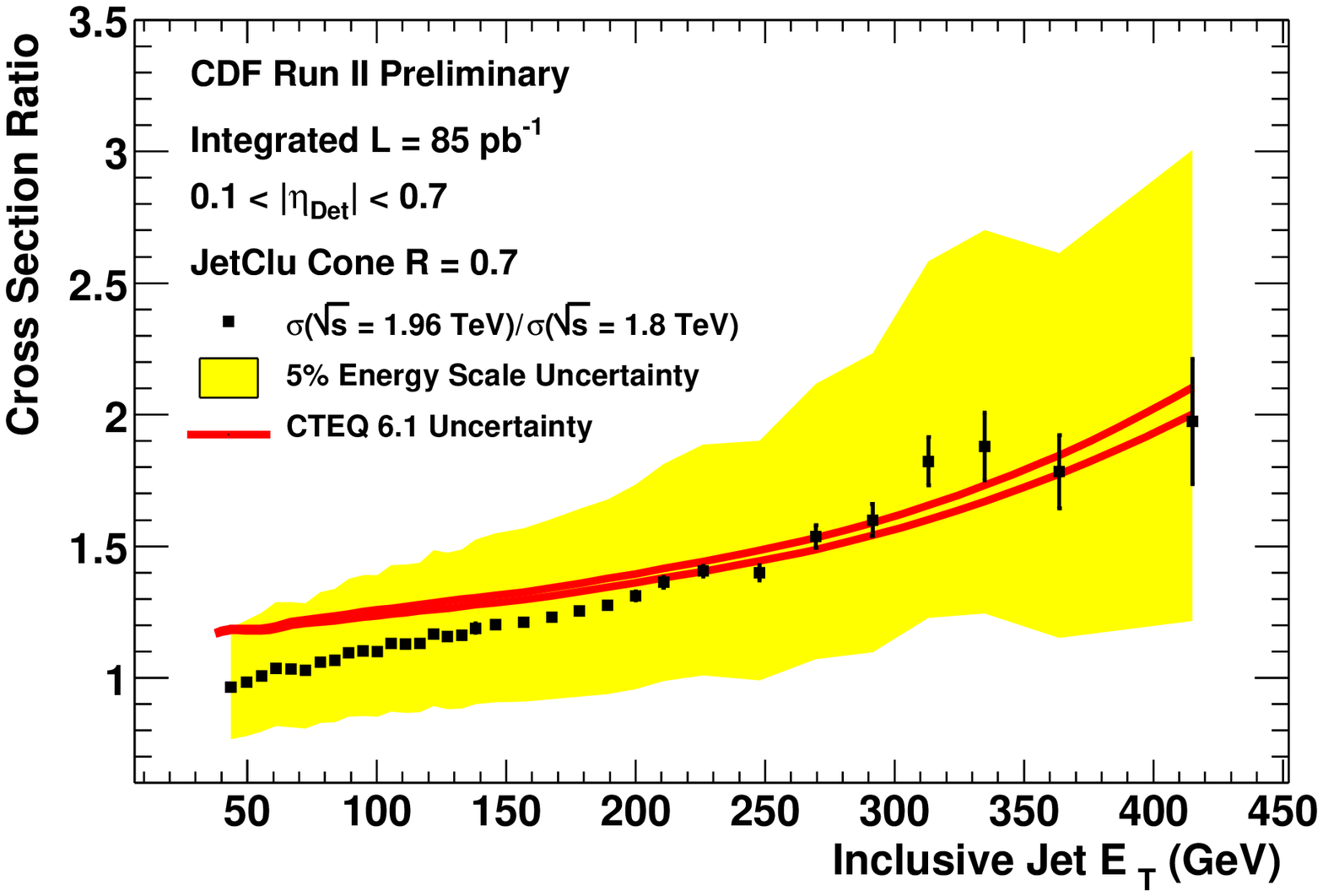,width=0.49\hsize}
\caption{
{\em (Left)} Run~II inclusive jet cross section (points) is compared to CTEQ~6.1 parton density functions. 
The band shows the change in cross section due to a 5\% jet energy scale shift;
{\em (right)} Ratio of Run~II/Run~I cross-sections compared to the QCD prediction using CTEQ6.1.
The lines show the uncertainty on CTEQ6.1.
\label{fig:xsec1}}
\end{figure}

Events were collected using different $E_T$ trigger thresholds with approriate prescale 
factors to balance the number of events in the different $E_T$ bins.
Data are used only when jet energies are well above the trigger thresholds and jets are fully efficient.
After accounting for the proper prescale factors, data are combined in a continuous $E_T$ spectrum, 
spanning more than eight orders of magnitude. 
The first preliminary measurement of the inclusive jet cross section is shown in Fig.~\ref{fig:xsec1}
and compared to the NLO QCD expectation determined using the CTEQ 6.1 parameterization~\cite{cteq} of the parton 
density functions. 
The fit reproduces well the distribution of the data points.
The data cover the $E_T$ range from 44 up to 550 GeV, extending the upper limit by
almost 150 GeV from Run~I.
The measured inclusive jet $E_T$ distribution is shown for the Run~I and Run~II data set. 
The jet cross section in Run~II is larger than that in Run~I due to the higher center-of-mass energy. 
This effect is especially prominent for high $E_T$ jets.

\section{Dijet Mass Distribution}

\begin{figure}
\epsfig{figure=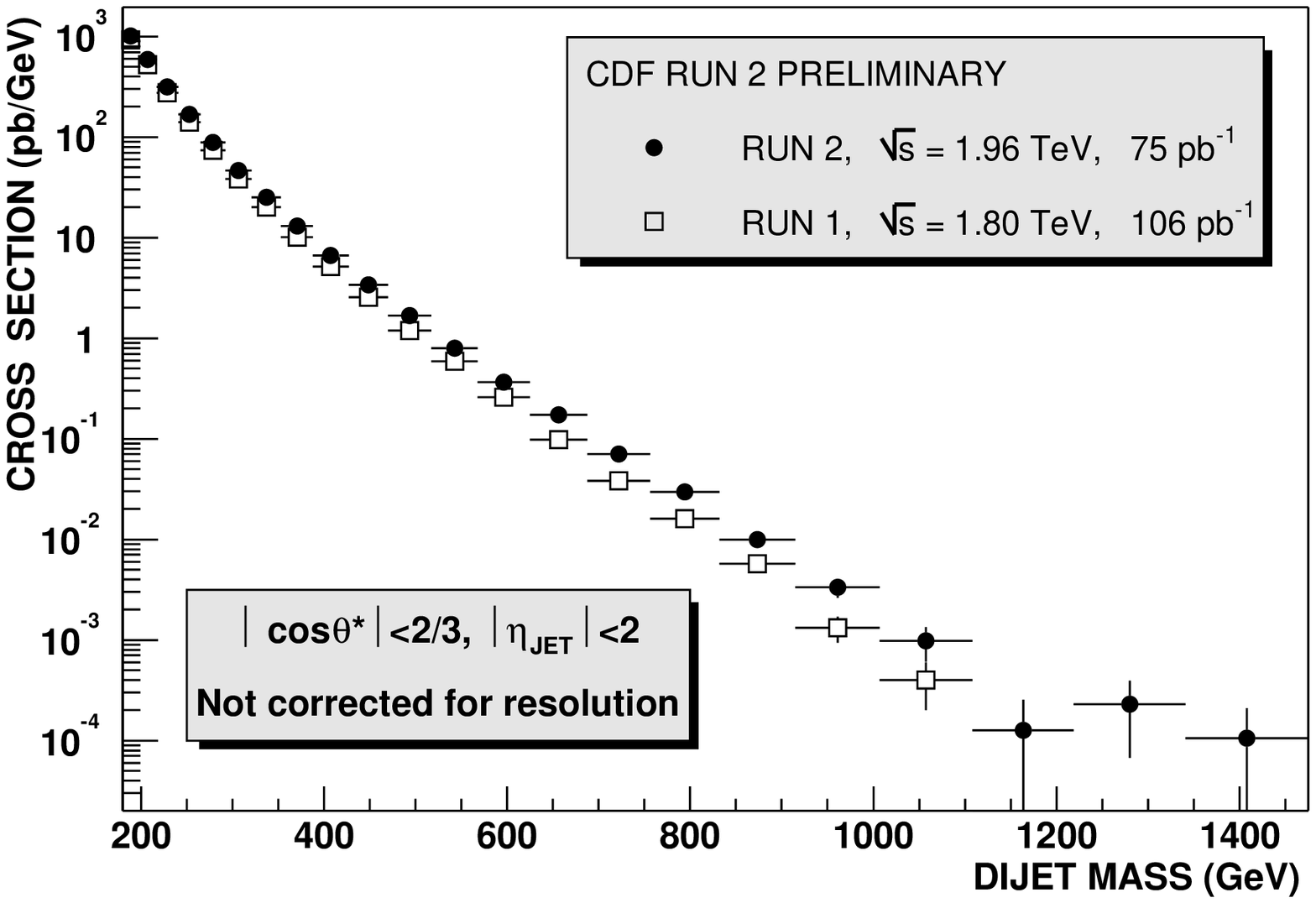,width=0.55\hsize}
\epsfig{figure=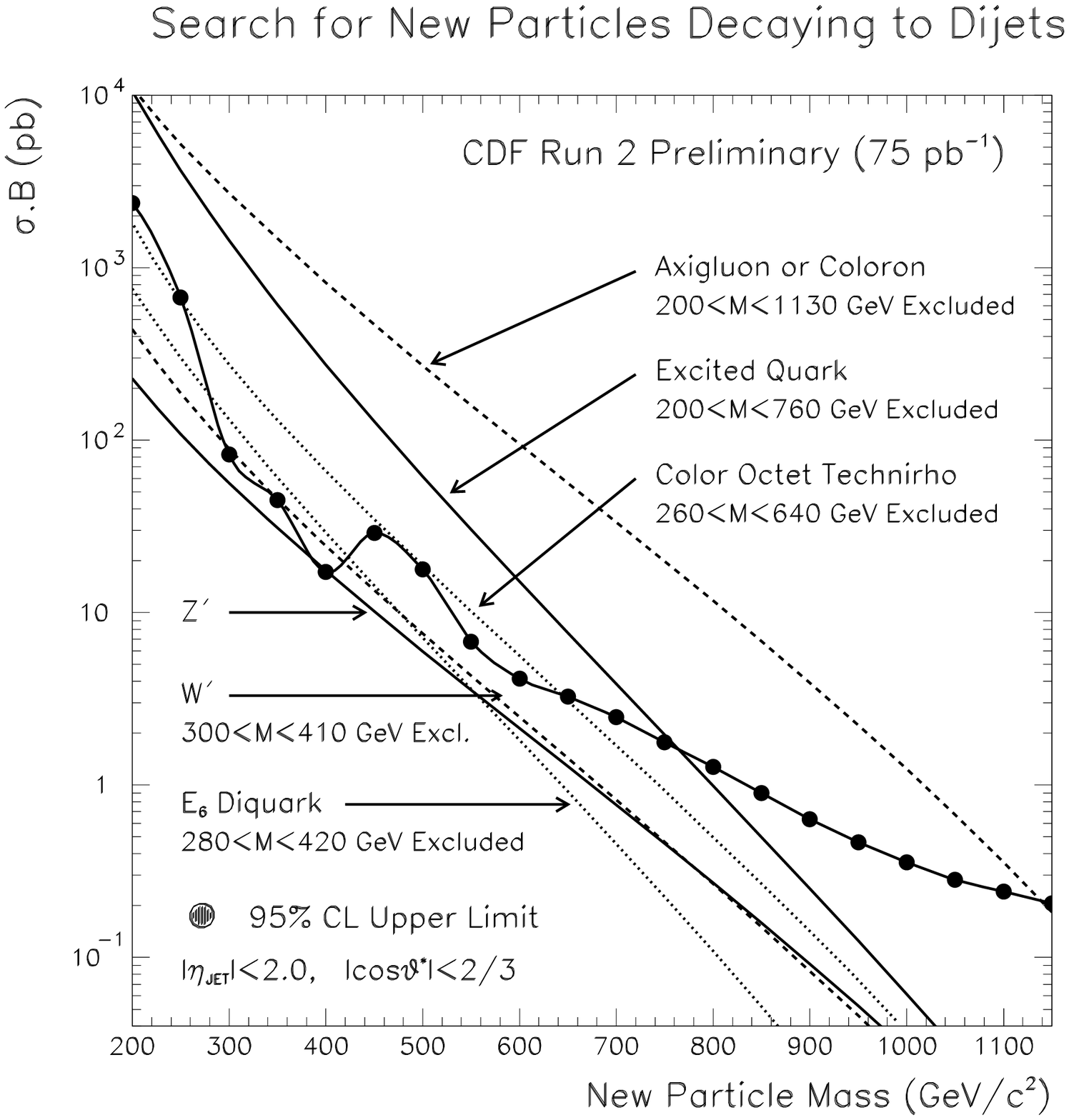,height=5.5cm}
\caption{{\em (Left)} Run~II dijet mass distribution (solid circles) compared to Run~I data (open boxes);
{\em (right)} The 95\% C.L. upper limit 
on the cross section times branching ratio for new particles 
decaying to dijets versus the new particles' mass (points) is compared to the theoretical predictions. 
The limit includes systematic uncertainties.
\label{fig:dijet}}
\end{figure}

Many classes of new particles can decay into two partons.
The large dijet sample available allows searching for new particles decaying into two jets with
a narrow resonance, significantly smaller than the measured dijet mass resolution.
The two jets with the highest transverse momentum in the event are used to define the dijet mass 
$m=\sqrt{(E_1+E_2)^2 + (\vec{\\\ p_1}+\vec{\\\ p_2})^2}$.
The differential cross section 
of the dijet mass spectrum is plotted versus the mean dijet mass (Fig.~\ref{fig:dijet}, left), where
the bin width approximately equals the dijet mass resolution. The distribution begins at 180 GeV and falls steeply up to
the highest mass event with a dijet mass of $\approx 1400$GeV. 
The Run~II cross section is everywhere above the Run~I spectrum.
The requirement of $|\cos\theta^\star|<2/3$ is used to reduce the QCD background, which peaks at $|\cos\theta^\star|=1$.

To search for new particles, a fit to the data points is performed with a simple background parameterization plus a narrow resonance.
A search for bumps comparable with our mass resolution can be achieved.
We set 95\% C.L. upper limits on the cross section times branching ratio for narrow dijet resonances and compare them with
predictions from various models (axigluons, flavor universal colorons, excited quarks, color octet technirhos, 
E6 diquarks, W', Z', etc.).
Data above the fit in more than two consecutive bins will mimic the shape of a new particle and increase 
the limit on the cross section. 
Around 230~GeV there are three consecutive points above the fit, and around 470~GeV there are two consecutive points above the fit.
These excesses can be observed in Fig.~\ref{fig:dijet} (right).
There is no significant evidence for new particles.

\section{Diffractive Dijet Events}
The diffractive dijet events studied are characterized by the
presence of two jets resulting from a hard scattering and a leading
antiproton, which escapes the collision intact. 
The Forward Detectors~\cite{lishep}, which are designed to enhance
the capabilities for studies of diffractive physics, 
include the {\it Roman Pot} (RP) fiber tracker to detect leading antiprotons,
a set of {\it Beam Shower Counters} (BSCs)
to tag rapidity gaps, and two forward MiniPlug calorimeters~\cite{mp_nim}.
In addition, the BSCs help vetoing events with multiple interactions.
The leading antiproton is scattered with a small momentum transfer and loses a 
fraction $\xi_{\overline{p}}$ of its initial momentum.
The associated dijets are detected in the Central and Plug calorimeters.

\begin{figure}
\epsfig{figure=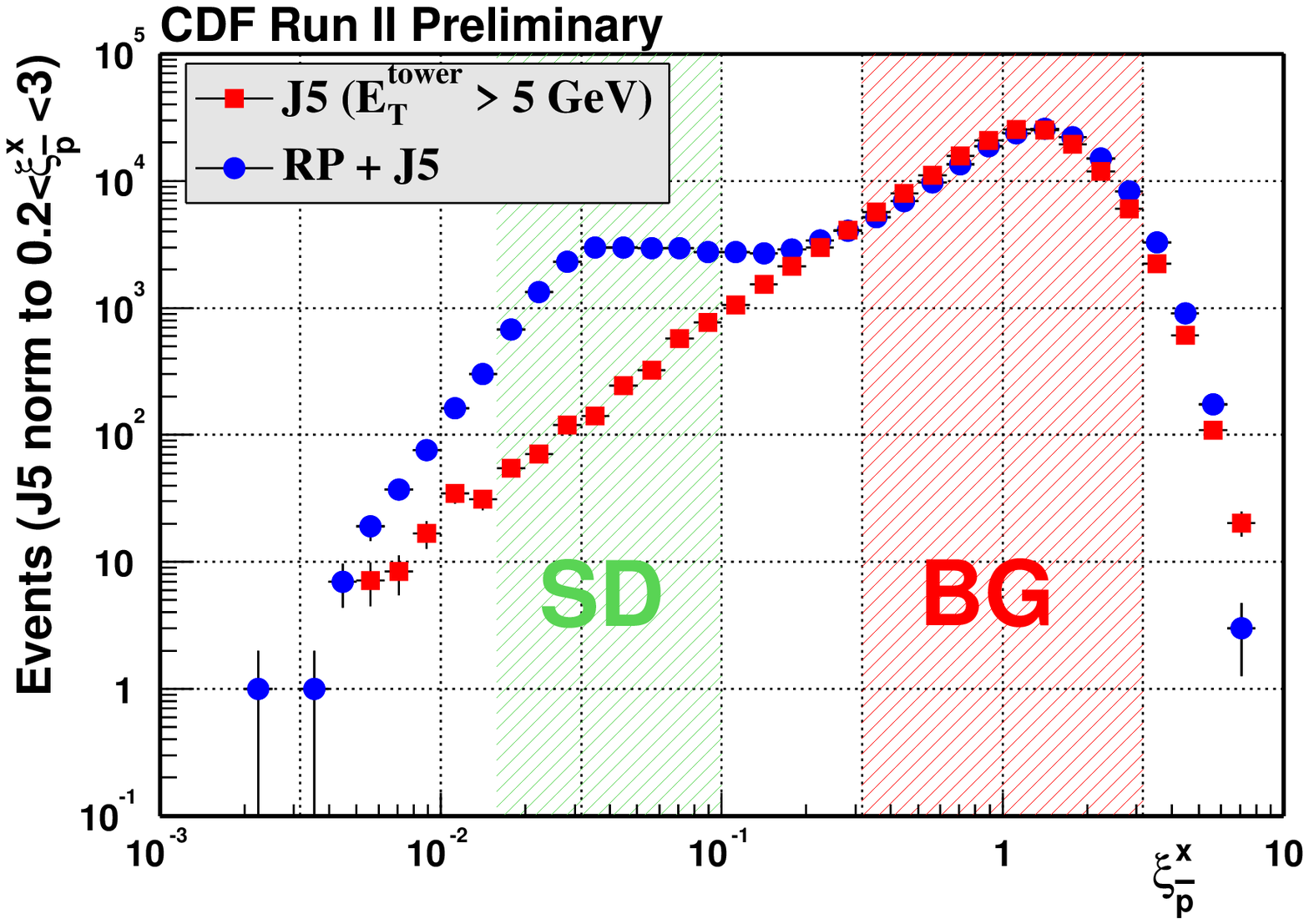,width=0.49\hsize}
\epsfig{figure=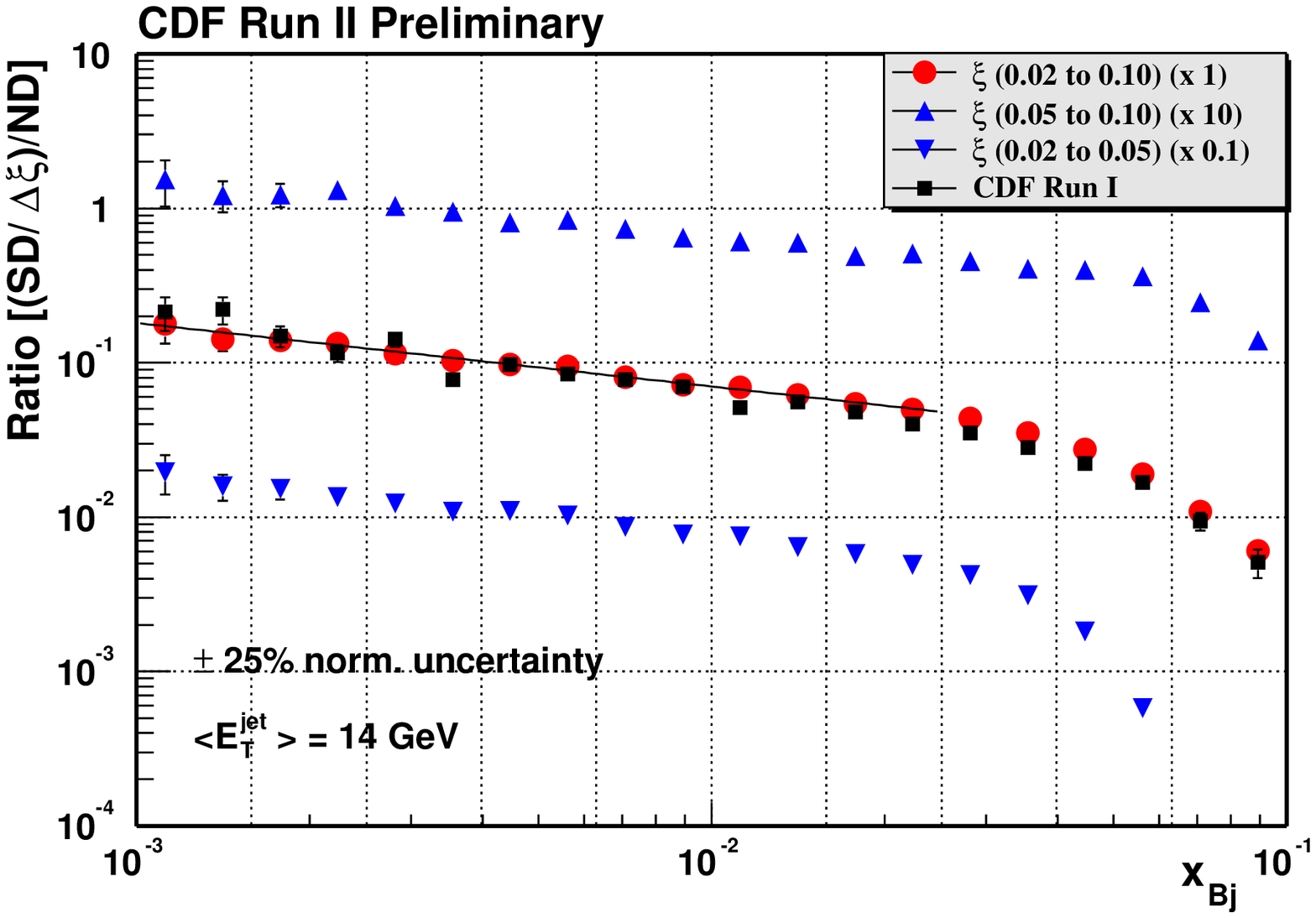,width=0.49\hsize}
\vspace*{-0.5cm}
\caption{{\em Left}: Momentum loss of the antiproton ($\xi_{\overline{p}}$) distribution in the data sample (circles) 
compared to a generic jet sample (squares).
SD and BG regions are selected according to the measured $\xi$ values;
{\em Right}: Ratio of diffractive to non-diffractive dijet event rates as a function of $x_{Bj}$ 
(momentum fraction of parton in the antiproton).
\label{fig:diffractive}}
\end{figure}

The full circles in Fig.~\ref{fig:diffractive} (left) indicate the $\xi_{\overline{p}}$ distribution of the entire data sample.
Diffractive events (SD) have a small $\xi_{\overline{p}}$ value, 
whereas non-diffractive events (ND) have $\xi_{\overline{p}}\approx 1$, as
the antiproton breaks up and loses entirely its initial momentum.
The smearing of the distribution around $\xi_{\overline{p}}\sim 1$ results mainly from the energy resolution measurement.
The shaded $\xi$ regions at small and large 
values indicate the chosen SD and ND sub-samples, respectively.
The ratio ($\cal{R}$) of SD to ND dijet production rates is studied as a function of the Bjorken scaling variable $x_{Bj}$
of the struck parton in the antiproton (Fig.~\ref{fig:diffractive}, right).
In Leading Order QCD, $\cal{R}$ is the ratio of the diffractive
to ND parton densities of the antiproton in dijet production.
The $x_{Bj}$ dependence of the production rate of SD events 
is steeper than that of the ND.
Results for the entire SD interval 
are consistent with Run~I data.
Results are also shown for two separate $\xi$ bins, 
showing no dependance on the chosen $\xi$ region used.
Furthermore, preliminary results indicate that the ratio does not depend on $E{_T}^2 \equiv Q^2 $, 
in the range from $Q^2 = 100$~GeV$^2$ up to 1600~GeV$^2$, 
indicating that the $Q^2$ evolution of the Pomeron is similar to that of the proton.

Dijet production of Double Pomeron Exchange (DPE) events\cite{dpe} occur when both 
proton and antiproton escape intact the collision.
These events are characterized by rapidity gaps on both sides of the interaction point.
Using a dedicated trigger, approximately 17,000 events have already been selected during Run~II 
($\sim$100 events were found in Run~I).
The jet $E_T$ distribution indicates a similar behavior for both SD and DPE events.
The exclusive dijet production rate of these DPE events will be of great interest to resolve the 
exclusive Higgs production debate~\cite{higgs}.

\vspace*{-.1cm}
\section{Conclusions}
\vspace*{-.2cm}
After one year of data-taking, the first physics results from Run~II presented here 
re-established the CDF measurements from Run~I and stepped beyond,
allowing optimism for a bright future of exciting measurements.

\vspace*{-.1cm}
\section*{Acknowledgments}
\vspace*{-.2cm}
Many thanks to the organizers of the Moriond conference and to all CDF collaborators
who worked hard for many years in order to accomplish the results presented here.

\end{document}